\documentclass[twocolumn,preprintnumbers,nofootinbib]{revtex4}
\usepackage{graphicx}
\usepackage{amssymb}
\usepackage{color}
\usepackage{enumerate}
\usepackage{amsmath}
\def\beq{\begin{equation}}
\def\eeq{\end{equation}}
\def\beqn{\begin{eqnarray}}
\def\eeqn{\end{eqnarray}}

\renewcommand{\texttt}{{}}
\newcommand{\be}{\begin{eqnarray}}
\newcommand{\ee}{\end{eqnarray}}

\begin{document}

\title{
{%
Multidimensional 
finite quantum 
gravity 
}} 
\author{Leonardo Modesto}
\email{lmodesto@fudan.edu.cn}

\affiliation{
{\small Department of Physics \& Center for Field Theory and Particle Physics,} \\
{\small Fudan University, 200433 Shanghai, China}
}

\date{\small\today}

\begin{abstract} \noindent
We advance a class of unitary higher derivative theories of gravity 
that realize an ultraviolet completion of Einstein general relativity in any dimension. 
This range of theories is marked by an  
entire function, which averts
extra degrees of freedom (including poltergeists) and improves the high energy
behavior of the loop amplitudes. 
It is proved that only one-loop divergences survive and the
theory can be made 
super-renormalizable regardless of the spacetime dimension. 
Moreover, using the Pauli-Villars regularization procedure introduced by 
Diaz-Troost-van Nieuwenhuizen-van Proeyen (DTPN)  
and applied to Einstein's gravity by Anselmi, we are able to remove
the divergences also at one-loop, making the theory completely finite in any dimension 
as expected by Anselmi and Asorey-Lopez-Shapiro.

\end{abstract}
\pacs{05.45.Df, 04.60.Pp}
\keywords{perturbative quantum gravity, nonlocal field theory}

\maketitle


\paragraph*{Introduction ---}
The microscopic structure of the universe is thoroughly compatible with 
quantum field theory and the standard model
of particle physics based on renormalizability and perturbative theory. In this paper
we assume the latter ones as the guiding principles for all fundamental interactions and we 
seek a new theory of gravity that can encompass such features. 
It is erroneously thought 
 that general relativity and quantum mechanics are not compatible, but there is nothing inconsistent between them. 
Just like for the Fermi theory of weak interactions, quantum Einstein's gravity is
perfectly consistent, solid and calculable, 
it is only non-renormalizable. 
At short distances, higher order 
operators in the Lagrangian become decisive 
suggesting that we need an ultraviolet completion of Einstein's gravity.
The question is: what are the relevant operators?
The answer lies in a ``{New Classical Theory of Gravity}" that is  
free of singularities at classical level and renormalizable or finite at quantum level. 
We believe that the two objectives are interdependent. 
Therefore, the aim of this work is to extend {\em classical} Einstein-Hilbert theory 
to make gravity compatible with the above guiding principles (renormalization and perturbative theory) 
in the ``quantum field theory framework". 
We begin with a new unitary higher derivative theory for gravity in a multidimensional spacetime
\cite{modesto, BM, M2, M3, M4, Krasnikov, Tombo,Khoury:2006fg} 
(see also \cite{calcagniNL, E1, E2, E3, E4, E5, Moffat1, Moffat2, Moffat3, corni1} and 
\cite{Deser1, Deser2, Modesto:2013jea, Maggiore} for nonlocal 
infrared extended theories of gravity), 
and we show that the quantum theory can be made super-renormalizable
because only one-loop divergences survive. 
Moreover, these one-loop divergences can be removed by introducing Pauli-Villars 
determinant regulators (in the number of $2 D \times 3$, where $D$ is the spacetime dimension) 
in the Diaz-Troost-van Nieuwenhuizen-van Proeyen formalism (DTPN) \cite{Diaz:1989nx} 
which explicitly preserves covariance.
Action, measure, and regularization procedure are BRS (Becchi-Rouet-Stora) invariant 
\cite{Anselmi:1991wb,Anselmi:1992hv, Anselmi:1993cu}. 
We end up with a completely
finite theory (
all the beta functions can consistently annulled) 
and the
outcome is a completely ``finite quantum field theory of gravity" in any dimension.  
This work is a completion of the previous work on polynomial \cite{shapiro} and non-polynomial 
\cite{modesto, BM, M2, M3, M4,
Krasnikov, Tombo}, 
super-renormalizable quantum gravity applying Anselmi's 
scheme 
\cite{Anselmi:1993cu}. 

\paragraph*{New gravity ---}
The aim of this section is to define a 
 ``new theory of gravity" 
in a $D$-dimensional spacetime assuming the following 
consistency requirements. 
\begin{enumerate}
\renewcommand{\theenumi}{\arabic{enumi}}
\item Unitarity. A general theory is well defined if
``tachyons" and ``ghosts" are absent, in which case the
corresponding propagator has only first poles 
with real masses (no tachyons) and with positive
residues (no ghosts.)
\item Super-renormalizability or finiteness. This
hypothesis makes consistent the theory at quantum level in analogy
with all the known fundamental interactions.
\item Lorentz invariance. 
This is a symmetry of
nature well tested experimentally beyond the Planck mass.
%
\item Last but not least, 
the energy conditions can not be violated on the matter side, 
but they can be violated on the gravity side because of the higher-derivative
operators in the classical theory. This property is crucial to avoid the ``singularities" that plague almost all the solutions of Einstein's gravity \cite{ModestoMoffatNico, CalcagniModesto, BambiMalaModesto1, BambiMalaModesto2, koshe1, koshe2, koshe3, koshe4, V1}. 
\end{enumerate}

The minimal action for gravity, which will prove compatible with the above properties, reads as follows\footnote{
{\em Definitions ---}
The metric tensor $g_{\mu \nu}$ has 
signature $(+ - \dots -)$ and the curvature tensors are defined as follows: 
$R^{\mu}_{\nu \rho \sigma} = - \partial_{\sigma} \Gamma^{\mu}_{\nu \rho} + \dots $, 
$R_{\mu \nu} = R^{\rho}_{\mu \rho \nu}$, 
$R = g^{\mu \nu} R_{\mu \nu}$},
\be
 \hspace{-0.0cm}
{S}_{\rm g} = - \!\!  \int \!\! d^D x \sqrt{ - g} \, 2 \kappa_D^{-2} \Big( R  
+ G_{\mu \nu} \,  \frac{ e^{H(-\Box_{\Lambda})} -1}{\Box}  \,  R^{\mu \nu} \Big) , 
\label{Action}
\ee
where $z \equiv - \Box_{\Lambda} \equiv - \Box/\Lambda^2$ and $\Box = g^{\mu\nu} \nabla_{\mu} \nabla_{\nu}$ is the covariant d'Alembertian operator. $\Lambda$ is an invariant fundamental mass scale and $G_{\mu\nu}$ is the Einstein's tensor. The entire function $V^{-1}(z) \equiv \exp H(z)$ 
($H(z)$ is in turn an entire function) is non-polynomial and 
satisfies the following general properties \cite{Tombo}:
\begin{enumerate}
\renewcommand{\theenumi}{(\roman{enumi})}
\item 
%
$V^{-1}(z)$ is real and positive on the real axis and it has no zeroes on the 
whole complex plane $|z| < + \infty$. This requirement assures the absence of 
extra gauge-invariant poles other than the transverse massless physical graviton pole.
A maximal extension of the theory (\ref{Action}) compatible with unitarity has been published in \cite{Briscese:2013lna}.
\item
$|V^{-1}(z)|$ has the same asymptotic behavior along the real axis at $\pm \infty$.
\item 
we define  
$2 \mathrm{N} + 4 = D_{\rm odd} +1$ in odd dimension and $2 \mathrm{N} + 4 = D_{\rm even}$
in even dimension to avoid fractional powers of the D'Alembertian operator. 
Therefore, it exists $\Theta>0$ such that 
\be
&& \lim_{|z|\rightarrow + \infty} |V^{-1}(z)| \rightarrow | z |^{\gamma + \mathrm{N}+1}, \nonumber \\
&& \gamma\geqslant D_{\rm even}/2 
\,\,\,\, {\rm and} 
\,\,\, \, 
 \gamma\geqslant (D_{\rm odd}-1)/2 \, , 
\label{tombocond}
\ee 
for the argument of $z$ in the following conical regions  
$C = \{ z \, | \,\, - \Theta < {\rm arg} z < + \Theta \, ,  
\,\,  \pi - \Theta < {\rm arg} z < \pi + \Theta\}$,
for $0< \Theta < \pi/2$. 
The last condition is necessary in order to achieve the highest convergence
of the theory in the ultraviolet regime. 
The necessary 
asymptotic behavior is imposed not only on the real axis, but also on the conic regions that surround it.  
In an Euclidean spacetime, the condition (ii) is not strictly necessary if (iii) applies.
\end{enumerate}

An explicit example of $\exp H(z)$ that satisfies the properties (i)-(iii) can be easily constructed \cite{Tombo}, 
\be
&& \hspace{-0.5cm}
V^{-1}(z) \equiv e^{H(z)}=
e^{\frac{1}{2} \left[ \Gamma \left(0, p^2(z) \right)+\gamma_E  \right] } \, \left| p(z) \right|  \nonumber \\
&&\hspace{-0.5cm}
=  \underbrace{
e^{\frac{\gamma_E}{2}} \,
\left| p(z) \right| 
}_{V^{-1}_{\infty}(z) } +
    \underbrace{ 
\left(   e^{\frac{1}{2}  \Gamma \left(0, p^2(z) \right)  }  -1 \right) 
e^{\frac{\gamma_E}{2}} 
\, \left| p(z) \right|  
}_{V^{-1}(z) -V^{-1}_{\infty}(z) \, \sim \, e^{- F(z)}  \,\,, \,\, \,\, F(z)>0}  , 
\label{Vlimit1}
\ee
where $p(z)$ is a polynomial of degree $\gamma +\mathrm{N}+1$ such that $p(0)=0$ and 
${\rm Re}( p^{2}(z) ) > 0$. 
In (\ref{Vlimit1})
 $\gamma_E=0.577216$ is the Euler's constant and 
$
\Gamma(a,z) = \int_z^{+ \infty} t^{a -1} e^{-t} d t
$ 
is the incomplete gamma function.  
If we choose $p(z) = z^{\gamma +N+ 1}$ 
the $\Theta$ angle defining 
the cone $C$ of (iii) turns out to be $\Theta = \pi/(4 \gamma +4 \mathrm{N} + 4)$. 
For $p(z) =z^{\gamma + \mathrm{N} + 1} \equiv z^n$ 
($n \equiv \gamma + \mathrm{N} + 1$),
the function $F(z)$ in (\ref{Vlimit1})
is well approximated, at least along the real axis, by $F(z) \propto z^m$ with 
$m \in \mathbb{N}$, $m \gtrsim n$.
Two examples are: 
$F(z) \approx 2 |z|^5$ for $n=4$ or 
$F(z) \approx 2 z^{12}$ for $n=10$. 
A crucial property of the form factor for the convergence of the theory is that 
%
\be
&& \hspace{-0.5cm} \lim_{z \rightarrow +\infty} V(z)^{-1} := V^{-1}_{\infty}(z) = e^{\frac{\gamma_E}{2}} \, |z|^{\gamma + \mathrm{N} +1} \,\,\,\, 
\,\,\,\, {\rm and} \nonumber \\ 
&& \hspace{-0.5cm} 
 \lim_{z \rightarrow +\infty} 
\left(\frac{V(z)^{-1}}{e^{\frac{\gamma_E}{2}} |z|^{\gamma + \mathrm{N} +1} } - 1 \right) z^n = 0
\,\,\,\, \forall \, n \in \mathbb{N}\, .
\label{property}
\ee
%
%
%
%
\paragraph*{Gauss-Bonnet operator ---}
In $D>4$ the Gauss-Bonnet invariant is not a total derivative; however, at the same time it does not affects the propagator around flat spacetime and it can only contribute to the interaction vertexes. In the theory (\ref{Action}) it does not even affect the divergent part of the one-loop effective action, but it gives a contribution to the finite parts of the one-loop diagrams.

\paragraph*{Propagator ---}
\label{gravitonpropagator}
%
Splitting the spacetime metric in the flat Minkowski background and the fluctuation $h_{\mu \nu}$ 
defined by $g_{\mu \nu} =  \eta_{\mu\nu} + \kappa_D \, h_{\mu\nu}$,
we can expand the action (\ref{Action}) to the second order in $h_{\mu\nu}$.
The result of this expansion together with the usual harmonic gauge fixing term reads \cite{HigherDG}:
$\mathcal{L}_{\rm lin} + \mathcal{L}_{\rm GF} = h^{\mu\nu} \mathcal{O}_{\mu\nu, \rho\sigma} \, h^{\rho\sigma}/2$, 
where the operator 
$\mathcal{O}$ is made of two terms, one coming from the linearization of (\ref{Action})
and the other from the following gauge-fixing term,
$\mathcal{L}_{\rm GF}  = \xi^{-1}  \partial^{\nu}h_{\nu \mu } \omega(-\Box_{\Lambda}) \partial_{\rho}h^{\rho \mu}$ 
($\omega( - \Box_{\Lambda})$ is a weight functional \cite{Stelle, Shapirobook, Tombo}.)
The d'Alembertian operator in $\mathcal{L}_{\rm lin}$ and in the functional weight $\omega$ must be conceived on the flat spacetime. 
Inverting the operator $\mathcal{O}$ \cite{HigherDG}, we find the 
two-point function in the harmonic gauge ($\partial^{\mu} h_{\mu\nu} = 0$),
\be
&& \hspace{-0.8cm}   \mathcal{O}^{-1} \!=\!
 \frac{V(  k^2/\Lambda^2 )  } {k^2} \!
\left( P^{(2)} 
- \frac{P^{(0)}}{D-2 }  \right) \!+ \!
 \frac{\xi (2P^{(1)} + \bar{P}^{(0)} ) }{2 k^2 \, \omega( k^2/\Lambda^2)} .
\ee
The tensorial 
indexes for the operator $\mathcal{O}^{-1}$ and the projectors $\{ P^{(0)},P^{(2)},P^{(1)},\bar{P}^{(0)}\}$ have been omitted. 
The above projectors are defined by 
\cite{HigherDG, VN}\label{proje2}: 
 $P^{(2)}_{\mu \nu, \rho \sigma}(k) = ( \theta_{\mu \rho} \theta_{\nu \sigma} +
 \theta_{\mu \sigma} \theta_{\nu \rho} )/2 - \theta_{\mu \nu} \theta_{\rho \sigma}/(D-1)$,
$P^{(1)}_{\mu \nu, \rho \sigma}(k) = \left( \theta_{\mu \rho} \omega_{\nu \sigma} +
 \theta_{\mu \sigma} \omega_{\nu \rho}  +
 \theta_{\nu \rho} \omega_{\mu \sigma}  +
  \theta_{\nu \sigma} \omega_{\mu \rho}  \right)/2$,  
$P^{(0)} _{\mu\nu, \rho\sigma} (k) = \theta_{\mu \nu} \theta_{\rho \sigma}/(D-1)$, 
$\bar{P}^{(0)} _{\mu\nu, \rho\sigma} (k) =  \omega_{\mu \nu} \omega_{\rho \sigma}$, 
 $\theta_{\mu \nu} = \eta_{\mu \nu} - k_{\mu } k_{\nu }/k^2$ and $\omega_{\mu \nu } = k_{\mu} k_{\nu}/k^2$.
%
%


\paragraph*{Power counting and super-renormalizability ---} 
Let us then examine the ultraviolet behavior of the quantum theory.
According to the property (iii) in the high energy regime, 
the propagator in the momentum space
and the leading interaction vertex are schematically given by
\be
&& \hspace{-0.5cm}
\mathcal{O}^{-1}(k) \sim \frac{1}{k^{2 \gamma +2 \mathrm{N} +4}} \,\,\,\,\,\,
\mbox{in the ultraviolet} \,,
\label{OV} \\
&& \hspace{-0.5cm}
{\mathcal V}^{(n+2)} \!
\sim  h^n \, \Box_{\eta} h   \frac{ V^{-1}( - \Box_{\Lambda})}{\Box}  \Box_{\eta} h
\, \rightarrow \,   h^n \, \Box_{\eta} h
\,   \Box_{\eta}
^{\gamma + \mathrm{N} } \,
\Box_{\eta} h\,,
\nonumber
\ee
where $\Box_{\eta} := \eta^{\mu\nu} \partial_{\mu} \partial_{\nu}$.
In (\ref{OV}) the indices for the gravitational fluctuations
$h_{\mu \nu}$ are omitted.
From (\ref{OV}), the upper bound to the superficial degree of divergence 
in a spacetime of ``even" or ``odd" dimension reads 
\be
&&
\omega(G)_{\rm even} =  D_{\rm even} - 2 \gamma  (L - 1) \, , 
\label{even}\\
&& \omega(G)_{\rm odd} = D_{\rm odd} - (2 \gamma+1)  (L - 1).
\label{odd}
\ee
In (\ref{even}) and (\ref{odd}) we used the topological relation between vertexes $V$, internal lines $I$ and 
number of loops $L$: $I = V + L -1$. 
Thus, if $\gamma > D_{\rm even}/2$ or $\gamma > (D_{\rm odd}-1)/2$, 
only 1-loop divergences survive in this theory. Therefore, 
the theory can be made super-renormalizable by introducing in the classical action local operators 
of mass dimensionality up to $M^D$
\cite{modesto, BM, M2, M3,M4, Krasnikov, E1, E2, E3, E4, E5}. 
{Once more let us reiterate what we mean by ``renormalizable theory".
A theory is renormalizable {\em iff} all the divergent contributions to the effective action are proportional to operators already present in the classical theory.}

%

\paragraph*{Quantum gravity ---}
In the previous sections we showed unitary around flat stacetime and power-counting super-renormalizability.
In this section we quantize the theory defined by (\ref{Action})
in the ``path-integral formulation". We use the background field method and we focus our attention 
on the DTPN-Pauli-Villars regularization \cite{Diaz:1989nx}, which preserves covariance: the regularized Lagrangian, the 
measure and the regularization procedure turn out to be BRS-invariant (quantum symmetry which involves
the graviton and the ghost fields after gauge fixing.)
This procedure permits  to remove the one-loop divergences without modifying the classical Lagrangian (\ref{Action}). Since multi-loop amplitudes are convergent, we get a completely finite theory of quantum gravity (all beta functions vanish).
By imposing appropriate and consistent conditions on the Pauli-Villars fields, we will be able to remove
all the divergences between the maximal ones having the form of a cosmological constant
and the logarithmic ones 
proportional to $R_{\mu\nu\rho\sigma}^{D/2}$. 

The Lagrangian (\ref{Action}) can be regularized with a set of complex vectors $W_{i \mu}$, a set of real vectors $Z_{i \mu}$, and a set of real tensor $T_{i \mu\nu}$. 
In the background field method 
the ghosts are regularized by the complex and the real vectors, while the metric is regularized by the real tensors. 
However, this is just a convention because in general $W,Z$ and $T$ altogether regularize the 
entire range of divergences \cite{Anselmi:1991wb}.
The regularized Lagrangian is made of six operators,
\be
 \hspace{-0.1cm}
\mathcal{L} = \mathcal{L}_{\rm g} +  \mathcal{L}_{\rm GF} + \mathcal{L}_{\rm GH} 
+\sum_{i=1}^{n_t} \mathcal{L}_{t_i} + 
\sum_{i=1}^{n_v} \mathcal{L}_{v_i} + \sum_{i=1}^{n_z} \mathcal{L}_{z_i} .
\label{FULL}
\ee
Since we are interested in the divergent contributions to the one-loop effective action,
the property (\ref{Vlimit1}) allows us to focus just on the ultraviolet  limit of (\ref{Action}),
namely
\be
\mathcal{L}_{\rm g} \sim - 2 \kappa^{-2}_D \sqrt{-g} \,  G_{\mu \nu}  \frac{ e^{\frac{\gamma_E}{2}} \, p(-\Box_{\Lambda}) }{\Box} R^{\mu \nu} .
\label{ActionUV}
\ee
The general polynomial involved in the divergent contributions to the one-loop effective action is
\be
\!\! 
 p_{\gamma+\mathrm{N}+1}(z) = a_{ \mathrm{N}} \, z^{\gamma + \mathrm{N} + 1} 
+ \dots + a_{\mathrm{N} - \frac{D}{2} } \, z^{\gamma + \mathrm{N}+1 - \frac{D}{2}}  .
\label{poly} 
\ee
The other operators coming from (\ref{Vlimit1}) can only contribute to the finite part because any correction to the leading ultraviolet limit is exponentially suppressed as shown by 
(\ref{property}).
$n_v$ is the number of $W$-vector fields, $n_z$ is the number of $Z$-vector fields and $n_t$ is the number of $T$-tensor fields, while 
\be
&&\hspace{-0.5cm}
 \mathcal{L}_{v_i} = \sqrt{-g} \Big[ \bar{W}_{i \mu} ( \Box + M^2_{v_i} ) W_{i}^{ \mu} \nonumber \\
 && \hspace{1.13cm}
 +  \bar{W}_{i \mu} ( \nabla^{\mu} \nabla_{\nu} - 2 \beta_g \nabla_{\nu} \nabla^{\mu} ) W_{i}^{ \nu} \Big]  \, ,
\label{W} 
\ee
\be
&&\hspace{-0.5cm}
 \mathcal{L}_{z_i} = \sqrt{-g} \Big[- \frac{1}{\alpha_g} {g}^{\mu \nu} \, Z_{i \mu} ( \Box + M^2_{z_i} ) 
( \Box + M^2_{z_i})_{\Lambda}^{  {\rm N} + \gamma} \, 
 Z_{i \nu} 
  \nonumber \\
&&\hspace{1.1cm}
 - \frac{1}{\alpha_g}
 Z_{i \mu} \, \left(\gamma_g {\nabla}^{\mu} {\nabla}^{\nu} - {\nabla}^{\nu} {\nabla}^{\mu} \right)  
 \Box_{\Lambda}^{  {\rm N} + \gamma} \, Z_{i \nu} \Big] \, , 
 \label{Z}
\\
&& \hspace{-0.5cm}
\mathcal{L}_{t_i} = \frac{2}{\kappa_D^{2}} \sqrt{-g} \Big[ \! - \frac{e^{\frac{\gamma_E}{2}}}{4} T_{i}^{\mu \nu} ( \Box + M^2_{t_i} )   p( - \Box - M_{t_i}^2  ) T_{i \mu \nu}  \nonumber \\
&&\hspace{-0.5cm}
+ T   R \, \nabla^{2 \rm{N} + 2 \gamma +2} T  + \dots + T   ( R \, \nabla^{D-4} R ) \nabla^{ 2 \gamma} T
\Big]  \, .
\label{T}
\ee
The constants $\alpha_g, \beta_g$ and $\gamma_g$ 
and more details about the action for the tensor fields $T_{i \mu \nu}$ will be given shortly. 
In the background field method the metric $g_{\mu\nu}$ splits in a background portion $\bar{g}_{\mu\nu}$ 
and in a quantum fluctuation $h_{\mu\nu}$,
\be
g_{\mu\nu} = \bar{g}_{\mu\nu} + h_{\mu \nu}. 
\label{BFM}
\ee
The non minimal couplings in the Lagrangians (\ref{W}), (\ref{Z}) and (\ref{T}) are chosen in order to look similar to $\mathcal{L}_{\rm GH}$ and 
 $\mathcal{L}_{\rm g}+ \mathcal{L}_{\rm GF}$, respectively, 
when both of them are 
expanded around the background $\bar{g}_{\mu\nu}$.
The gauge fixing and FP-ghost actions are  
\be
&& \hspace{-0.4cm} 
S_{\rm GF}  = \int d^D x \sqrt{ - \bar{g} } \, \chi_{\mu} \, C^{\mu \nu} \, \chi_{\nu} \, ,   
\,\,\, 
\chi_{\mu} = \bar{\nabla}_{\sigma} h^{\sigma}_{\mu} - \beta_g  \bar{\nabla}_{\mu} h \, , \nonumber \\
&&\hspace{-0.4cm} 
C^{\mu \nu} \! = - \frac{1}{\alpha_g}
 \left( \bar{g}^{\mu \nu} \Box 
 + \gamma_g \bar{\nabla}^{\mu} \bar{\nabla}^{\nu} - \bar{\nabla}^{\nu} \bar{\nabla}^{\mu} \right)  
 \Box_{\Lambda}^{  {\rm N} + \gamma}  , \nonumber \\
&& \hspace{-0.4cm} 
S_{\rm GH} = \int d^D x \sqrt{ - \bar{g} } \left[  \bar{C}_{\alpha} \, M^{\alpha}_{\beta} \, C^{\beta} 
+ b_{\alpha} C^{\alpha \beta} b_{\beta} \right] 
\, , \nonumber \\
&& \hspace{-0.4cm} 
M^{\alpha}_{\beta} = \Box \delta^{\alpha}_{\beta}  
+ \bar{\nabla}^{\alpha} \bar{\nabla}_{\beta} - 2 \beta_g \bar{\nabla}_{\beta} \bar{\nabla}^{\alpha}. 
\label{shapiro3a}
\ee
In (\ref{shapiro3a}) 
we used a 
gauge fixing with weight function $C^{\mu\nu}$  \cite{shapiro}. 
$\alpha_g, \beta_g, \gamma_g$ are dimensionless gauge fixing parameters, but 
the beta-functions are independent from them 
(see proof in \cite{shapiro}.) For $\beta_g =1/2$ and $\gamma_g =1$, 
the Lagrangians for the Pauli-Villars fields
(\ref{W}) and (\ref{Z}) and the Lagrangians for the ghosts $\bar{C},C,b$ simplify to
\be
&&\hspace{-0.45cm}
 \mathcal{L}_{C} \! =  \! \sqrt{-\bar{g} } \Big[ \bar{C}_{\mu} \Box  C_{i}^{ \mu}
 -  R^{\mu}\hspace{-0.02cm}_{\nu} \bar{C}_{\mu}  C^{ \nu} \Big]   ,
\label{CC}\\
&&\hspace{-0.45cm}
 \mathcal{L}_{v_i}  \! =  \! \sqrt{- \bar{g} } \Big[ \bar{W}_{i \mu} ( \Box + M^2_{v_i} ) W_{i}^{ \mu} 
 -  R^{\mu}\hspace{-0.02cm}_{\nu} \bar{W}_{i \mu}  W_{i}^{ \nu} \Big]   ,
\nonumber \\ 
 &&\hspace{-0.45cm}
 \mathcal{L}_{b}  \! =  \!  \frac{\sqrt{- \bar{g} }}{- \alpha_g}  \Big[ 
 b_{ \mu} \,
 \Lambda^2 
\Box _{\Lambda}^{  {\rm N}^{\prime}+1} 
 b^{ \mu} 
 -  b_{ \mu} \, R^{\mu \nu } 
 \Box_{\Lambda}^{  {\rm N}^{\prime}} \, b_{ \nu} \Big]  ,
 \label{bb}\\
&&\hspace{-0.45cm}
 \mathcal{L}_{z_i} \! =  \! \frac{\sqrt{- \bar{g}  }}{- \alpha_g}  \Big[ 
 Z_{i \mu} \Lambda^2 
( \Box + M^2_{z_i})_{\Lambda}^{  {\rm N}^{\prime}+1} 
 Z_{i}^{ \mu} 
 -  Z_{i \mu} \, R^{\mu \nu } 
 \Box_{\Lambda}^{  {\rm N}^{\prime} } \, Z_{i \nu} \Big]  ,
 \nonumber 
\ee
where ${\rm N}^{\prime} = {\rm N} + \gamma$. 
If we introduce new variables rescaling the fields $X \in  \{ \bar{W},W,Z,\bar{C},C,b \}$ by
$X \rightarrow (-g)^{-1/4} X$, we see that the mass terms for the Pauli-Villars fields 
give contribution to the propagator and to extra vertexes that are proportional to the mass square elevated to some integer power. 
The latter vertexes are present neither in the graviton, nor in the ghosts fields; however, 
they do not originate divergences, if the Pauli-Villars regularization conditions that will 
tackle later in the paper
(conditions (\ref{cConstraint}) and (\ref{cond3})) are satisfied. 
It follows that, in the redefined variables, $\mathcal{L}_{v_i}$ regularizes $\mathcal{L}_{C}$
and $\mathcal{L}_{z_i}$ regularizes $\mathcal{L}_{b}$. A similar analysis can be carried out for the
graviton and the tensors $T$. 

We have now all the conditions to define the partition function with the right functional measure 
compatible with BRS invariance \cite{Anselmi:1993cu},
\be
&& \hspace{-0.5cm}
Z = \int \! 
\mu(g, \bar{g} )
\prod_{\mu \leqslant \nu} \mathcal{D}g_{\mu\nu} \prod_{\nu} \mathcal{D} \bar{C}_{\nu}
 \prod_{\mu}  \mathcal{D}{C}^{\mu}
 \prod_{\mu}  \mathcal{D}{b}_{\mu}
 \nonumber \\
 && \hspace{-0.5cm}
 \prod_{i=1}^{n_v}  \left[   \prod_{\nu}  \mathcal{D}^{\prime} \bar{W}_{i \nu}
 \prod_{\mu}  \mathcal{D}^{\prime} W_{i}^{ \mu} 
 \right]  
  \prod_{i=1}^{n_z} \prod_{\mu}  \mathcal{D}^{\prime} Z_{i \mu} 
  \prod_{j=1}^{n_t}  \prod_{\mu \leqslant \nu}  \mathcal{D}^{\prime} T_{j \mu \nu}\nonumber \\
  && \hspace{-0.5cm}
 e^{i \int d^Dx \left[  \mathcal{L}_{\rm g} +  \mathcal{L}_{\rm GF} + \mathcal{L}_{\rm GH} + 
\sum_{i=1}^{n_v} \mathcal{L}_{v_i} + \sum_{i=1}^{n_z} \mathcal{L}_{z_i} 
+\sum_{i=1}^{n_t} \mathcal{L}_{t_i}
  \right]} \,  . 
  \label{QG}
 \ee
 We can evaluate the functional integral and express the partition function as a product of determinants,
 namely\footnote{The DTNP formalism is based on a particular definition of the functional integration 
 on the Pauli-Villars regulators. If $\chi_j$ is a bosonic Pauli-Villars field we have
 \be
 \int \mathcal{D}^{\prime} 
 \chi_j \, e^{\chi_j^{\rm T} A \chi_j} = ({\rm det} A)^{\frac{c_j}{2}} \,\,\,\, (\mbox{formal gaussian integral}) \, ,
 \ee
 where $A$ is a generic $\chi_j$-independent infinite matrix and $^{\rm T}$ denotes the transposition,
 while $c_j$ is a coefficient associated with the Pauli-Villars field $\chi_j$. 
 }
 \be
 && \hspace{-0.4cm}
  Z = e^{i S_{\rm g}(\bar{g}_{\mu \nu})}  \left\{ {\rm det} \! 
  \left[  \frac{\delta^2 (S_{\rm g} + S_{\rm GF}) }{\delta h_{\mu \nu} \delta h_{\rho \sigma}} \right]  \right\}^{-\frac{1}{2}}
 ({\rm det} \, M^{\alpha}_{\beta} ) 
\nonumber \\
&&\hspace{-0.4cm}
({\rm det} \, C^{\alpha \beta} )^{\frac{1}{2}}  \,\, 
\prod_{i=1}^{n_z} \!  \Bigg[ {\rm det} \Bigg( \underbrace{\frac{\delta S^2_{z_i} }{\delta Z_{i \mu} \delta Z_{i \nu} }}_{\mathcal{O}_{z_i}}     \Bigg)  \Bigg]^{\frac{c_{z_i}}{2}}
\!
\label{det}  \\
 &&   \hspace{-0.4cm}
 \prod_{i=1}^{n_v} \!  \Bigg[ {\rm det} \Bigg( \underbrace{\frac{\delta S^2_{v_i} }{\delta \bar{W}_{i \mu} \delta W_{i \nu} }}_{\mathcal{O}_{v_i}}     \Bigg)  \Bigg]^{\frac{c_{v_i}}{2}}
\!
 \prod_{i=1}^{n_t} \! \Bigg[ {\rm det} \Bigg( \underbrace{\frac{\delta S^2_{t_i} }{\delta T_{i \mu \nu} \delta T_{i \rho \sigma}}}_{\mathcal{O}_{t_i}} \Bigg)\Bigg]^{\frac{c_{t_i}}{2}} 
 \hspace{-0.2cm} , 
 \nonumber 
 \ee
 where $c_{v_i}$, $c_{z_i}$ and $c_{t_i}$ come from the definition of the functional integration on the Pauli-Villars 
 fields $W$, $Z$, $T$. Furthermore, we impose on them the following regularizations conditions 
 \cite{Anselmi:1993cu}
 \be
 \hspace{0.0cm}
&& \hspace{1cm} \boxed{ \sum_{j =1}^{n} c_{u_j} = c_u} \label{cConst0} \\ 
 && \hspace{-0.65cm}\boxed{ \sum_{j =1}^{n} c_{u_j} (M^2_j)^q= 0 \, , 
 \,\,\,\,  0 < q \leqslant \left[ \frac{D}{2} \right]}  
 \label{cConstraint} 
 \ee
 in which $[x]$ is the integer part of the number $x$.
 The index $u = v,z, t$ and the constants $c_v = -2$, $c_z = - 1$, $c_t = 1$.
%
To calculate the one-loop effective action we need to expand the action plus the gauge-fixing term to the second oder in the quantum fluctuation $h_{\mu \nu}$
\be
 H_{\mu \nu , \rho \sigma}= \frac{\delta^2 S_{\rm g}}{\delta h_{\rho \sigma} \delta h_{\mu \nu}}\Bigg|_{h=0} \!\!\!\!\!\!
 + \frac{\delta \chi_{\delta}}{\delta h_{\rho \sigma}} \, C^{\delta \tau } \, 
 \frac{\delta \chi_{\tau}}{\delta h_{\mu \nu}}\Bigg|_{h=0} 
 \!\!\!\!\!\!\! .
\ee
The explicit calculation of $H$ goes behind the scope of this paper and here we only offer  
the tensorial structure in terms of the curvature tensor of the background metric and its covariant derivatives. Since we are only interesting in the divergent contributions to the one-loop effective action,
from (\ref{ActionUV}) the {\em minimal} operator $H_{\mu \nu, \rho \sigma}$ consists only of 
the terms coming from the asymptotic limit of the form factor, namely 
%
%
%
\be
&& \hspace{-0.4cm} 
H= \Box^{\rm{N} + \gamma + 2} + 
(\bar{R} + \dots) \cdot
\bar{\nabla}^{2 \rm{N}+2 \gamma +2} \nonumber \\
&&\hspace{-0.4cm} 
+ 
(\bar{\nabla} \bar{R}  + \dots) \cdot \bar{\nabla}^{2 N+2 \gamma +1}
+  (\bar{R}^2 + \dots) \bar{\nabla}^{2 \rm{N} + 2 \gamma}
\nonumber \\
&& \hspace{-0.4cm} 
+(\bar{\nabla} \bar{R}^2 + \dots) \cdot \bar{\nabla}^{2 \rm{N}  +2 \gamma- 1}
 + (\bar{\nabla}^2 \bar{R}^2 + \dots) \cdot  \bar{\nabla}^{2 \rm{N} + 2 \gamma -2} 
\nonumber \\
&& \hspace{-0.4cm} 
+ \dots 
+ ( \bar{R} \, \bar{\nabla}^{D-4}   \bar{R} + \dots) \cdot \bar{\nabla}^{2 \rm{N} + 2 \gamma +4 - D} ,
%
\label{expantion}
\ee
where $\bar{\nabla}^Y \equiv \bar{\nabla}_{\mu_1} \dots \bar{\nabla}_{\mu_Y}$.
In (\ref{expantion}) we explicitly showed how the coefficients depend on the curvature tensors in a compact notation.
The dots inside the round brackets indicate operators with less derivatives of the background metric
tensor. These operators 
come from lower powers in the polynomial $p(z)$. 
For example, the last coefficient in (\ref{expantion}) reads,
\be
&& \hspace{-1cm}
  \bar{R} \, \bar{\nabla}^{D-4}   \bar{R} + \dots =  \bar{R} \, \bar{\nabla}^{D-4}   \bar{R} +  \bar{R} \, \bar{\nabla}^{D-6}   \bar{R} \nonumber \\
&& \hspace{1.8cm} + \dots +  \bar{R}^2  +  \bar{R} + {\rm const.} 
\ee
However, from here on 
we assume the polynomial in (\ref{poly}) to be proportional to $z^{\gamma +{\rm N}+1}$ 
to simplify our analysis. 

It is now clear how to define $\mathcal{L}_{t_i}$ in (\ref{T}) to regularize the gravitational Lagrangian, 
\be
\mathcal{L}_{t_i} =  \sqrt{-{g} } \, T_{i}^{\mu \nu}  \left(H_{\mu\nu, \rho \sigma}({g}_{\mu\nu})\Big|_{\Box \rightarrow \Box+M^2_{t_i}} \, \right)  T_{i}^{ \rho\sigma} \,  ,
\ee
where we replaced the background metric $\bar{g}_{\mu\nu}$ with ${g}_{\mu\nu}$ and all tensors are 
now defined by the metric $g_{\mu \nu}$. 
In the background field method the Lagrangians for $W$ and $T$ have also to be expanded using 
(\ref{BFM}). However, they are already quadratic in the Pauli-Villars fields and we can simply substitute $g_{\mu\nu}$ with $\bar{g}_{\mu\nu}$ in $\mathcal{L}_{v_i}$,
$\mathcal{L}_{z_i}$  and $\mathcal{L}_{t_i}$. 
The field redefinition $X \rightarrow (-g)^{-1/4} X$ also applies to the fields $T$. 
For the tensors $T$ as well as for the vectors $Z$ there are extra vertexes proportional to the mass,
which are absent for the graviton and the ghost $b$. However, such operators do not give divergent contributions to the one-loop amplitudes because of the set of conditions (\ref{cConstraint}) and (\ref{cond3}). 
The similarities between $\mathcal{L}_{\rm GH}$ and $\mathcal{L}_{v_i}+\mathcal{L}_{z_i} $,
and 
$\mathcal{L}_{\rm g} + \mathcal{L}_{\rm GF}$ and $\mathcal{L}_{t_i}$
are now evident since $\sum_i \mathcal{L}_{t_i}$ regularizes 
$\mathcal{L}_{\rm g} + \mathcal{L}_{\rm GF}$, while 
$\sum_i (\mathcal{L}_{v_i}+\mathcal{L}_{z_i})$ regularizes 
$\mathcal{L}_{\rm GH}$. 

We can now move on to calculate the one-loop effective action \cite{shapiro},
\be
&& \hspace{-1.1cm}
\Gamma^{(1)} = - i  \log Z= S_{\rm g}(\bar{g}_{\mu \nu}) 
\nonumber \\
&& 
+ \frac{i}{2}  \log {\rm det}(H) - i  \sum_{j=1}^{n_{t}} \frac{c_{t_j}}{2} \log {\rm det} ( \mathcal{O}_{t_j} )
\nonumber \\
&& 
- i \log {\rm det}(M)
- i  \sum_{j=1}^{n_{v}} \frac{c_{v_j}}{2} \log {\rm det} ( \mathcal{O}_{v_j} )\nonumber \\
&&
 - \frac{i}{2} \log {\rm det}( C )
 - i  \sum_{j=1}^{n_{z}} \frac{c_{z_j}}{2} \log {\rm det} ( \mathcal{O}_{z_j} ). 
\label{Gamma1}
\ee
Given a general operator $\mathcal{O}= \mathcal{O}_{0} + \mathcal{O}_{\rm I}$, 
where $\mathcal{O}_{0}$ refers to the free part and $\mathcal{O}_{\rm I}$ refers to the 
interaction part, 
\be
&&\hspace{-0.5cm}
 \log {\rm \det{\mathcal{O}}} = {\rm Tr} \log \mathcal{O} = 
{c} \, {\rm Tr} [ \log ( 1+ \mathcal{O}_{0}^{-1} \mathcal{O}_{\rm I} ) ]   \label{traccia}\\
&&\hspace{-0.5cm}
 = {c}  \sum_{n=1}^{+ \infty } \frac{(-1)^{n+1} }{n} \!\!   \int  \!\!
 d^D x_1 \dots d^Dx_n  \,
\mathcal{O}^{-1}_{0}(x_1 - x_2) \mathcal{O}_{\rm I}(x_2) \nonumber \\
&& \hspace{0.32cm}
\mathcal{O}^{-1}_{0}(x_2 - x_3) \mathcal{O}_{\rm I}(x_3)
\dots\dots \mathcal{O}^{-1}_{0}(x_{n} - x_1) \mathcal{O}_{\rm I}(x_1),
\nonumber 
\ee
where $c= {\rm constant}$.
In the background field method we define the fields $f_{\mu \nu}$ related to the background metric by 
$\bar{g}_{\mu \nu} = \eta_{\mu \nu} + f_{\mu \nu}$. 
$\mathcal{O}^{-1}_{0}$ ($O_{\rm I}$) is the propagator (interaction vertex) 
around flat spacetime for any field
circulating in the loop diagram: the graviton $h_{\mu \nu}$, the tensors $T_{i \mu\nu}$,
the ghosts $\bar{C}_{\nu}, C^{\mu},b_{\mu}$ or the vectors $W_{i \mu},\bar{W}_{i \nu}, Z_{i \mu}$. 

Let us now elaborate on the extra vertexes englobing the mass 
of the $Z$ vectors or $T$ tensors
when we expand the background metric around the flat spacetime.
For the case of the vectors $Z$, let us expand the action to the linear order in the fluctuation $f$ 
(a similar analysis applies to the tensors $T$),
\be
&& \mathcal{L}_{z_i} =  
 Z_{i \mu} \Lambda^2 \Big[ 
( \Box_0 + M^2_{z_i})_{\Lambda}^{  {\rm N}^{\prime}+1}  
\label{ZE}
\\
&& 
+ \sum_{k=1}^{{\rm N}^{\prime} + 1} \, ( \Box_0 + M^2_{z_i})^{k-1}  \, \delta \Box \, 
( \Box_0 + M^2_{z_i})^{{\rm N}^{\prime} + 1-k}
\Big] 
 Z_{i}^{ \mu}  \nonumber \\
 && -  Z_{i \mu} \, R^{\mu \nu } 
 \Box_{\Lambda}^{  {\rm N}^{\prime} } \, Z_{i \nu}  \nonumber \\
 && = 
 Z_{i \mu} \Lambda^2 \Big[ 
( \Box_0 + M^2_{z_i})_{\Lambda}^{  {\rm N}^{\prime}+1} 
+ {V}_1(
\partial^{ 2 {\rm N}^{\prime} +2 }, f)  \nonumber \\
&&
 +\underbrace{ {V}_2( 
 \partial^{ 2 {\rm N}^{\prime}  }, M_{z_i}^2, f) 
 + O(M^4_{z_i})}_{V_2(M_{z_i}^2, \dots, M_{z_i}^{2 \rm{N}^{\prime}} )}
\Big] 
 Z_{i}^{ \mu} -  Z_{i \mu} \, \underbrace{R^{\mu \nu } 
 \Box_{\Lambda}^{  {\rm N}^{\prime} }}_{V_R} \, Z_{i \nu} \, , 
 \nonumber 
\ee
where $\Box_0 = \eta^{\mu \nu}\partial_{\mu} \partial_{\nu}$. 
The vertexes ${V}_2$
are present only for the Pauli-Villars fields,
while the vertexes 
${V}_1$ are present for the $b$ ghost, too. 
However, the contribution of $V_2$ is zero 
because of the sets of constraints (\ref{cConstraint}) 
and 
(\ref{cond3}) 
(see the Appendix for more technical details\footnote{
We can put the determinant for the $Z$ fields in the following form, 
\be
 && \hspace{-0.4cm} 
  {\rm det}  \Big(( \Box_0 + M^2_{z_i})_{\Lambda}^{  {\rm N}^{\prime}+1}  \!
+ {V}_1 + V_2(M_{z_i}^2, \dots, M_{z_i}^{2 \rm{N}^{\prime}} ) 
+ V_R\Big)^{\!\! \frac{c_{z_j}}{2}}
 \nonumber 
 \\
 &&  \hspace{-0.4cm} 
 = {\rm det} \Big( ( \Box_0 + M^2_{z_i})_{\Lambda}^{  {\rm N}^{\prime}+1} \!
+ {V}_1 + V_R\Big)^{\!\! \frac{c_{z_j}}{2}}  \nonumber \\
&&\hspace{-0.4cm} 
\hspace{0.13cm} 
\cdot \, 
\underbrace{{\rm det} \left( 1 + \frac{V_2(M_{z_i}^2, \dots, M_{z_i}^{2 \rm{N}^{\prime}} ) + V_R}{( \Box_0 + M^2_{z_i})_{\Lambda}^{  {\rm N}^{\prime}+1} 
+ {V}_1 + V_R }  \right)^{\!\! \frac{c_{z_j}}{2}}}_{\hspace{-1.15cm} \lim_{ M_{z_i} \rightarrow + \infty } ( \,\dots \,  ) = 1} .
\ee
Taking the limit $M_{z_i} \rightarrow + \infty$ only the first factor allows the correct normalization
to regularize the $b$ ghost's action,
while the second factor tends to one. 
}.)

Using the background field method and the Pauli-Villars regularization, 
the main divergent integrals contributing to 
the one-loop effective action have the following form, 
\be
&& \hspace{-0.375cm} \includegraphics[height=2.8cm]{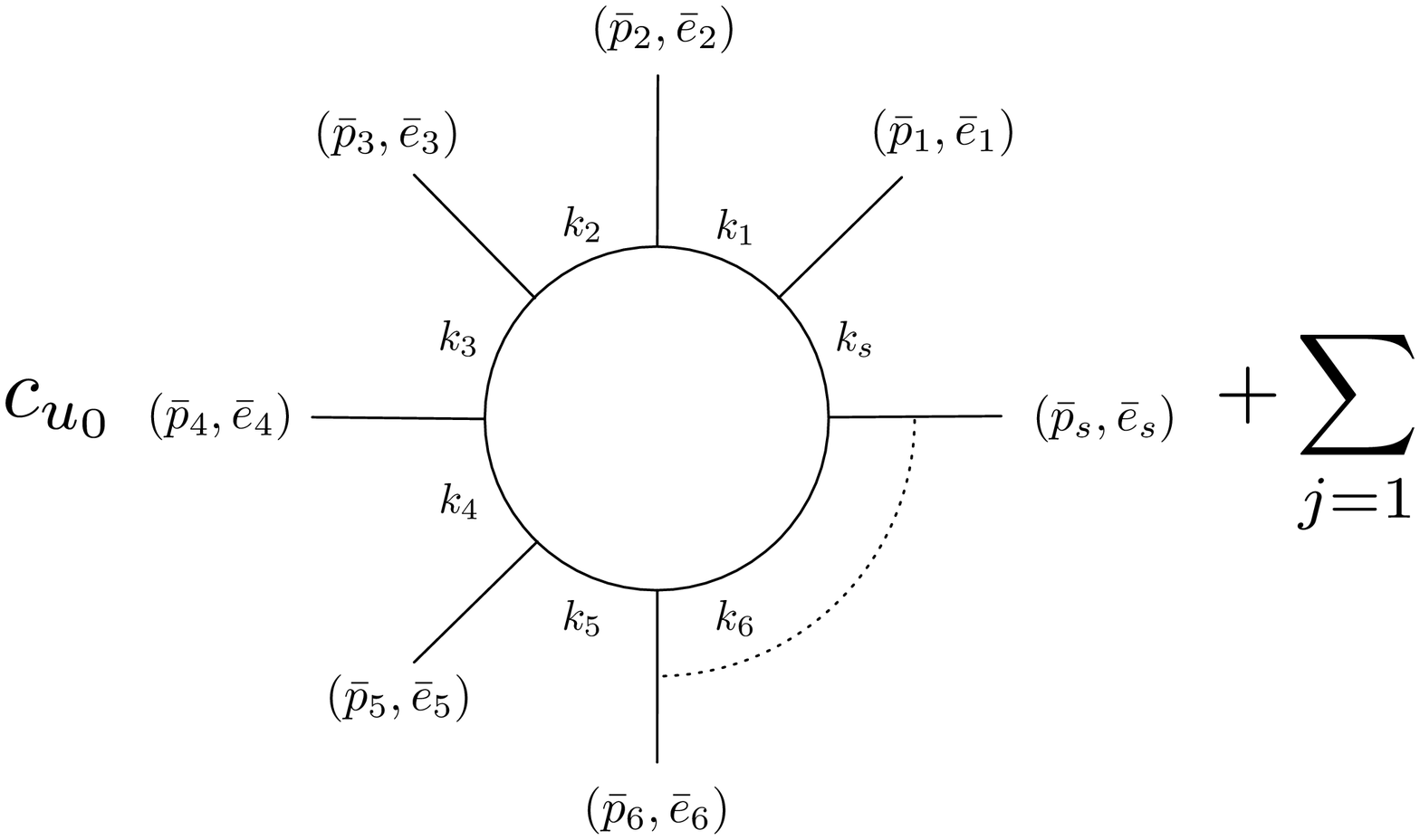} \nonumber\\
&& 
\hspace{-0.3cm}
\int \! \frac{d^D k}{(2 \pi)^D} {\sum_{j=0}}^{\prime}  c_{u_j} \left\{ \prod_{i=1}^{s} 
\frac{1}{[(k+p_i)^2 + M_{u_j}^2]^{n}} \right\} \! P(k)_{2 s  n}  . 
\label{integral}
\ee
$P_{2sn}(k)$ is a polynomial function of degree $2 n s$ in the momentum $k$ 
(generally it also relies on the external momenta $\bar{p}_a$), 
$p_i = \sum_{a=1}^i \bar{p}_a$, 
and $u=v,z,t$ labels the three different Pauli-Villars fields. 
The positive integer $n$ is: $n = \gamma +{\rm N} +2$ for $h_{\mu\nu}$ and $T_{i \mu \nu}$,
$n = \gamma +{\rm N} +1$ for $b_{\mu}$ and $Z_{i \mu}$, while 
$n = 1$ for $C,\bar{C}$ and $W,\bar{W}$.
The ${\sum}^{\prime}$ is from $j=0$ to $j=n_{u_j}$ and it also includes the massless fields with the convention $M_{u_0}=0$. $c_{v_0} =2$ for the ghosts $C,\bar{C}$ loops, $c_{z_0} =1$ for the ghost $b$ loops, and $c_{t_0} =-1$ for the case of the graviton loops. 
We can write, as usual,
\be
&& 
\prod_{i=1}^{s} 
\frac{1}{[
(k+p_i)^2 + M_{u_j}^2]^{n}}  = \nonumber\\
&& = {\rm c} \!  \int_0^1 \! \left( \prod_{i=1}^s x_i^{n-1} d x_i \!  \right) 
\delta\left( 1 - {\sum_{i=1}^s x_i} \right)
 \frac{1}{[k^{\prime 2} + M_{u_j}^{\prime 2}]^{n s}}  \, , \nonumber\\
 &&  
 k^{\prime} = k + \! \sum_{i=1}^s x_i p_i  \, , \,\, \nonumber \\
 && 
 M^{\prime 2}_{u_{j}} = M^{2}_{u_{j}} - \left( \sum_{i=1}^s x_i p_i \right)^{\!\! 2} \!\! + 
 \sum_{i=1}^s p_i^2 x_i  \, .
\ee
where ${\rm c} = {\rm constant}$. 
In (\ref{integral}), we move outside the convergent integral in $x_i$ and we replace 
$k^{\prime}$ with $k$ 
\be
\int \!\! \frac{d^D k}{(2 \pi)^D} {\sum_{j=0}}^{\prime}  c_{u_j} 
\frac{P^{\prime}(k, p_i, x_i)_{2 s  n} }{( k^2 + M_{u_j}^{\prime 2})^{n s }} \, .
\label{int2}
\ee
Using Lorentz invariance and missing the argument $x_i$, we replace the polynomial 
$P^{\prime}(k, p_i, x_i)_{2 n s}$ with a polynomial of degree $n \times s$ in $k^2$, 
namely $P^{\prime \prime}(k^2, p_i)_{n s}$.
In the intermediate steps we integrate (\ref{int2}) 
in $|k|$ from zero to a cut-off $\Lambda_{\rm c}$ and then
we take the limit $\Lambda_{\rm c} \rightarrow \infty$.  
Therefore the integral (\ref{int2}) reduces to
\be
\int_0^{\Lambda_{\rm c}} \!\!\!  \frac{d^D k}{(2 \pi)^D} {\sum_{j=0}}^{\prime}  c_{u_j} 
\frac{P^{\prime \prime}(k^2, p_i)_{n s} }{( k^2 + M_{u_j}^{\prime 2} )^{n s }} \, .
\label{int3}
\ee
We can decompose the polynomial $P^{\prime \prime}(k^2, p_i)_{n s} $ 
in a pro-duct of external and internal momenta
only to obtain the divergent contributions,
\be
&& \hspace{-0.4cm} 
P^{\prime \prime}(k^2, p_i)_{n s} = \sum_{\ell=0}^{[ D/2 ]} \alpha_{\ell}(p_i) k^{2 n s-2 \ell} = \nonumber \\
&& \hspace{-0.4cm} 
k^{2 n s} \alpha_0 + k^{2 n s-2} \alpha_1(p_i) + k^{2 n s - 4 } \alpha_2(p_i)   
+ \dots \, . 
\ee
By changing variables to $y = |k|^2/M^{\prime 2}_{u_j}$ and  neglecting multiplicative constants common to all the fields, the integral (\ref{int3}) turns out to be,
\be
 && \hspace{-0.5cm}
 \mathcal{I} = {\sum_{j=0}}^{\prime}  c_{u_j} \!\!
\int_0^{\frac{\Lambda^2_{\rm c}}{M_{u_j}^{\prime 2}}   } \! d y \,
y^{\frac{D-2}{2}}
\frac{M_{u_j}^{\prime D}}{(1+y)^{n s}}
\Big[  \alpha_0 \, y^{n s}  + 
\label{int4} 
 \\
&& \hspace{-0.5cm}
+ \frac{\alpha_1(p_i)}{ M_{u_j}^{\prime 2} } \, y^{n s -1} + 
\frac{\alpha_2(p_i)}{ M_{u_j}^{\prime 4} } \, y^{n s -2}
+ 
\frac{\alpha_3(p_i)}{ M_{u_j}^{\prime 6} } \, y^{n s -3}
 + \dots 
\Big]  \nonumber   \\
&& \hspace{-0.5cm}
= {\sum_{j=0}}^{\prime}  c_{u_j} \!\!
\int_0^{\frac{\Lambda^2_{\rm c}}{M_{u_j}^{\prime 2}}   } \! d y \,
y^{\frac{D-2}{2}}
\frac{M_{u_j}^{\prime D}}{(1+y)^{n s}}
   \! \sum_{\ell = 0}^{[ D/2 ]} \frac{\alpha_{\ell} (p_i)}{ M_{u_j}^{\prime 2 \ell } } y^{n s - \ell} \, .
\nonumber 
\ee
The integral is straightforward and the result for general $D$ 
is given in the appendix. 
Finally, the conditions we must add to (\ref{cConst0}) and (\ref{cConstraint}) 
for the integrals (\ref{int4}) (or (\ref{int5})) to vanish 
are 
\be
\boxed{\,\, \sum_j c_{u_j} (M_{u_j}^2)^p \left(  \log \frac{M^2_{u_j}}{\mu_u^2}  \right)^{\epsilon_D} = 0 \,\, }
\label{cond3}
 \ee
where $\epsilon_D = 1$ and $p = 0,1,2, \dots, D/2$ for $D$ even,
while $\epsilon_D = 0$ and $p = 1/2,3/2, \dots, D/2$ for $D$ odd. 
In this way the theory is fully regularized. 
The price to pay to make the theory finite is the introduction of three arbitrary constants $\mu_u$ \cite{Anselmi:1993cu}.

\paragraph*{Finite quantum gravity in even and odd dimension ---}
Using the results of the above section we can now show that all the beta-functions are zero 
and the theory is finite in any dimension.  
Suppose the masses of the Pauli-Villars fields are taken to be finite, for now, and consider the terms of
the one-loop effective action that are expected to diverge when one lets the Pauli-Villars
masses go to infinity. 
This terms in even dimension can only be proportional to 
\be
&&
\hspace{-0.0cm}
\Lambda_u^D =  \sum_{j=1} c_{u_j} M^D_{u_j} \log \frac{M_{u_j}^2}{\mu_u}\, , \,\,  \nonumber \\
&& 
\Lambda_u^{D-2} =  \sum_{j=1} c_{u_j} M^{D-2}_{u_j} \log \frac{M_{u_j}^2}{\mu_u}\, , \,\, \nonumber \\
&& 
\Lambda_u^{D-4} =  \sum_{j=1} c_{u_j} M^{D-4}_{u_j} \log \frac{M_{u_j}^2}{\mu_u}\, , \,\, \nonumber \\
&& 
\Lambda_u^{D-6} =  \sum_{j=1} c_{u_j} M^{D-6}_{u_j} \log \frac{M_{u_j}^2}{\mu_u}\, , \,\, \nonumber \\
&& \dots \nonumber \\
&& \dots \nonumber \\
&& \Lambda_u^{2} =  \sum_{j=1} c_{u_j} M^{2}_{u_j} \log \frac{M_{u_j}^2}{\mu_u}\, , \,\, \nonumber \\
&& \log \frac{{\Lambda_u^2}}{\mu_u}=  \sum_{j=1} c_{u_j} \log \frac{M_{u_j}^2}{\mu_u}\, . \,\, 
\ee
The other divergent contributions are absent because the properties (\ref{cConst0}) and (\ref{cConstraint})
of the Pauli-Villars fields have been enforced. 
The effective regularized Lagrangian 
including only divergent contributions reads 
\be
&&  \hspace{-0.6cm}
\boxed{\mathcal{L}_{\rm eff}
=  - \sqrt{|g|} \, 2 \kappa_D^{-2} \left( R + G_{\mu \nu} \,  \frac{ e^{H(-\Box_{\Lambda})} -1}{\Box}  \,  R^{\mu \nu} \right)} \nonumber \\
&& \hspace{-0.6cm}
+\sum_{u}\sum_{j=1} c_{u_j} M^D_{u_j} \log \frac{M_{u_j}^2}{\mu_u} \, a_u \sqrt{|g|} \nonumber \\
&& \hspace{-0.6cm}
+ \sum_{u}  \sum_{j=1} c_{u_j} M^{D-2}_{u_j} \log \frac{M_{u_j}^2}{\mu_u} \, b_u \sqrt{|g|}  R\nonumber\\
&& \hspace{-0.6cm}
+  \sum_{u}  \sum_{j=1} c_{u_j} M^{D-4}_{u_j} \log \frac{M_{u_j}^2}{\mu_u} \sqrt{|g|} ( 
c_u^{(1)}  R^2+ c_u^{(2)} R_{\mu\nu}^2 
\nonumber \\
&& \hspace{-0.6cm}
+  c_u^{(3)} R_{\mu\nu\rho\sigma} R^{\mu\nu\rho\sigma} ) 
\nonumber \\
&& \hspace{-0.6cm}
+ \, \dots\dots \nonumber \\
&& \hspace{-0.6cm}
+ \sum_{u} \sum_{j=1} c_{u_j} \log \frac{M_{u_j}^2}{\mu_u} 
\sqrt{-g} 
\Big( d_u^{(1)} R^{{\rm N}+2}_{\dots}   \nonumber \\
&& \hspace{-0.6cm}
+ d_u^{(2)} R^{{\rm N}-1}_{\dots} \nabla R_{\dots} \nabla R_{\dots} 
+d_u^{(3)} R_{\dots} \Box^{\rm N} R_{\dots} +\dots\Big)  \,  .
\label{counterterms}
\ee
When the conditions in (\ref{cond3}) are applied, all the above operators 
from the second row onwards vanish 
and 
the classical Lagrangian (\ref{Action}), highlighted with a box in (\ref{counterterms}), turns out to be completely regularized.

\paragraph*{Explicit evaluation of the integrals (\ref{integral}) ---} 
We hereby explicitly calculate the integral (\ref{int4}) in a multidimensional spacetime: 
\be
&&  \hspace{-0.2cm}
\mathcal{I} = 
{\sum_{j=0}}^{\prime}  c_{u_j} M_{u_j}^{\prime D}
 \sum_{\ell = 0}^{[ D/2 ]} \frac{\alpha_{\ell} (p_i)}{ M_{u_j}^{\prime 2 \ell } } 
e^{-\frac{1}{2} i \pi  (D-2 \ell+2 n s)} \nonumber \\
&& \hspace{0.5cm}
{\rm B}_{- (  \Lambda_{\rm c}^2/M_{u_j}^{\prime 2 })^2  }\left(\frac{D}{2} - \ell +n s,1-n s\right) ,
  \nonumber 
\ee
where $B_z(a,b) = \int_0^z t^{a-1} (1-t)^{b-1} dt$. 
Expanding the result for large values of the cut-off $\Lambda_{\rm c}$ we find 
\be
&& \hspace{-0.2cm}
\mathcal{I} 
= {\sum_{j=0}}^{\prime}  c_{u_j} M_{u_j}^{\prime D}
 \sum_{\ell = 0}^{[ D/2 ]} \frac{\alpha_{\ell} (p_i)}{ M_{u_j}^{\prime 2 {\ell }} } \times 
\nonumber \\
&& \hspace{-0.2cm}
\frac{1}{\Gamma (n s)} \Bigg[ \Gamma \left(\ell -\frac{D}{2}\right) 
\Gamma \!
   \left(\frac{D}{2}- \ell +n s  \right) 
   \nonumber 
\\
   &&\hspace{-0.2cm}
+ \left( \frac{\Lambda_{\rm c}^2}{M_{u_j}^{\prime 2 }} \right)^{\!\! D-2  \ell }  
\Bigg(\frac{2 \Gamma (n s)}{D-2 \ell} 
+\frac{2 \Gamma (n s+1)}{(-D+2 \ell+2)
   \left(\frac{\Lambda_{\rm c}^2}{M_{u_j}^{\prime 2 }} \right)^{\!\! 2}} \nonumber \\
   &&\hspace{-0.2cm}
   +\frac{\Gamma (n s+2)}{(d-2 (\ell+2)) \left(\frac{\Lambda_{\rm c}^2}{M_{u_j}^{\prime 2 }} \right)^{\! 4}}
   +\frac{\Gamma (n s+3)}{( 6 (\ell +3)-3 D ) \left(\frac{\Lambda_{\rm c}^2}{M_{u_j}^{\prime 2 }} \right)^{\! 6}}
   \nonumber \\
   &&\hspace{-0.2cm} 
   + \, O  \!  \left(\frac{M_{u_j}^{\prime 2 }}{\Lambda_{\rm c}^2}\right)^{\!\! 4}  \Bigg)  \Bigg] , 
   \nonumber 
   \ee
 where again $[x]$ is the integer part of the number $x$.
 
   Finally, we explicitly evaluate the sum on the integer $\ell$, namely
   \be
   && \hspace{0.0cm}
   \mathcal{I} 
= {\sum_{j=0}}^{\prime}  c_{u_j} M_{u_j}^{\prime D}
  \Bigg\{ \alpha_0 \Bigg[ a_D^{(1)} \left(\frac{\Lambda_{\rm c}   }{  M_{u_j}^{\prime  }}\right)^{\!\! D} 
  \nonumber \\
  && \hspace{0cm} 
  + a_{D-2}^{(1)} \left(\frac{\Lambda_{\rm c}   }{  M_{u_j}^{\prime  }}\right)^{\!\! D-2} 
  +\dots + a_1^{(1)}
  \log \left(\frac{\Lambda_{\rm c}   }{  M_{u_j}^{\prime  }}\right)^{\!\! 2} 
  + a_0^{(1)} \Bigg]
  \nonumber \\
  &&  \hspace{0cm}
  +
  \frac{\alpha_1(p_i)}{M_{u_j}^{\prime  2 }} \Bigg[ 
   a_{D-2}^{(2)} \left(\frac{\Lambda_{\rm c}   }{  M_{u_j}^{\prime  }}\right)^{\!\! D-2} 
  + a_{D-4}^{(2)} \left(\frac{\Lambda_{\rm c}   }{  M_{u_j}^{\prime  }}\right)^{\!\! D-4}  \nonumber \\
  &&\hspace{0cm}
   +\dots + a_1^{(2)}
  \log \left(\frac{\Lambda_{\rm c}   }{  M_{u_j}^{\prime  }}\right)^{\!\! 2} 
  + a_0^{(2)} \Bigg] + \dots  \nonumber \\
 &&  \hspace{0cm}
  +
  \frac{\alpha_{[ D/2 ]}(p_i)}{M_{u_j}^{\prime  D }} \Bigg[ a_1^{([D/2])}
     \log \left(\frac{\Lambda_{\rm c}   }{  M_{u_j}^{\prime  }}\right)^{\!\! 2} + a_0^{([D/2])} \Bigg] \Bigg\}.
     \label{int5}
   \ee
In this last expression (\ref{int5}), 
$a_i^{(j)}$ 
are numerical constants depending on $n$ and/or $s$, while the logarithmic contributions must be understood only in even dimension. 

For the vertexes $V_2$ 
in (\ref{ZE}) (analog vertexes are present for the $T$ tensors), 
the polynomial 
in (\ref{integral}) reads $P(k)_{2 s  n-2}$ because it has at least two less derivatives and the result 
in (\ref{int5}) will be proportional to at least one power of $M_{z_j}^{2}$ or $M_{t_j}^{2}$ ($j\geqslant1$.)
Therefore, we will only need the relation (\ref{cConstraint}) to make the divergent integrals vanish.
This paragraph makes clear why we do not have to worry about the extra vertexes, 
although there is no similar contribution coming from the ghost $b$ (or the graviton $h$ for the case 
of the tensors $T$.)

\begin{figure}
\begin{center}
\includegraphics[height=2.8cm]{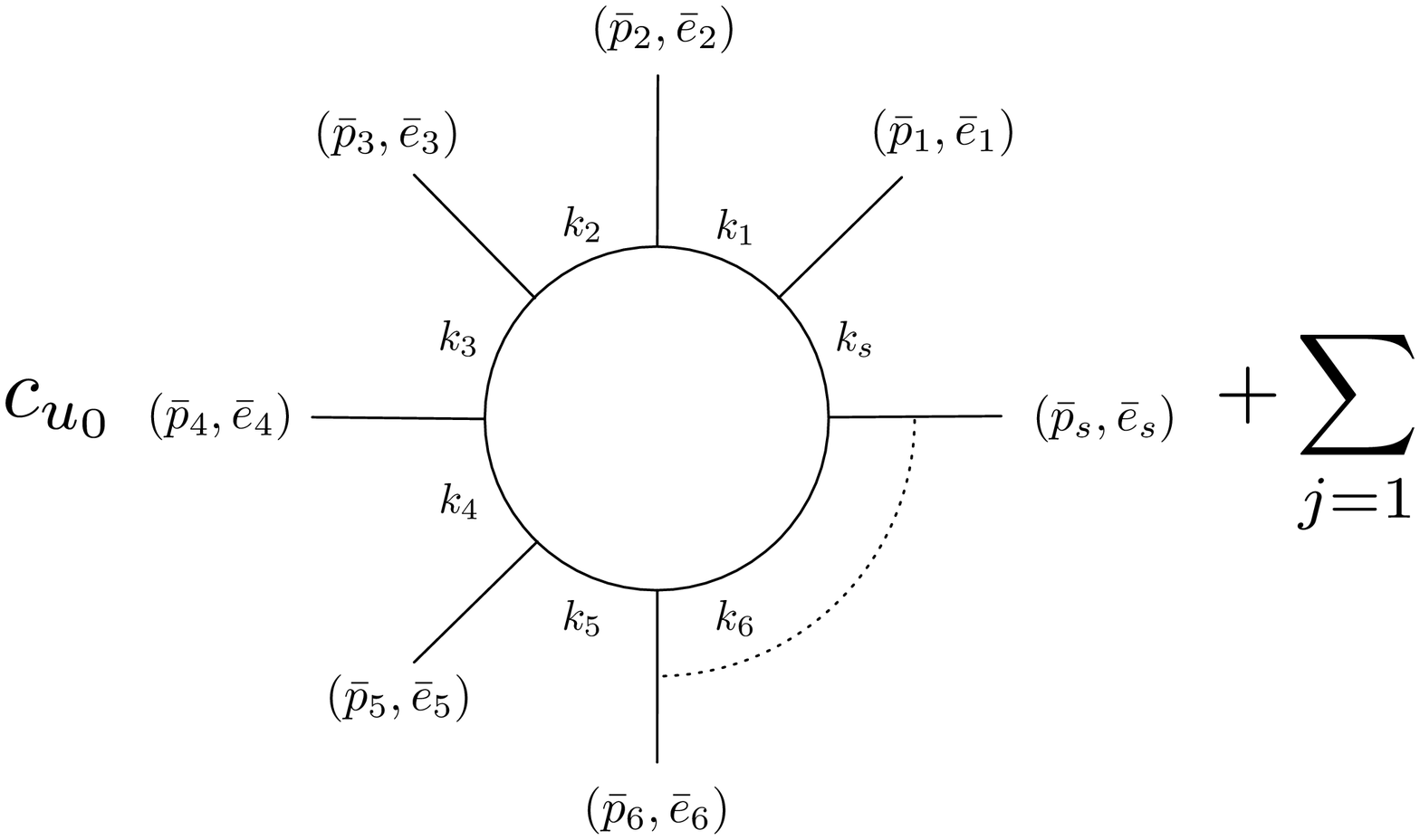}
\end{center}
\caption{\label{fig1} Diagrammatic 
representation of the one-loop cancellation due to
the Pauli-Villars fields. The fields $h_{\mu \nu}$, $C^{\mu},\bar{C}_{\nu}$ and $b_{\alpha}$ circulate in
the loop on the left, while the fields $T_{i \mu \nu}$, $W_i^{\mu},\bar{W}_{i \nu}$ and $Z_{i \mu}$ circulate in the loop on the right. $u_0$ and $u_j$ are defined in the text. }
\end{figure}

\paragraph*{Conclusions ---} 
In this paper we explicitly showed that a class of unitary non-polynomial theories of gravity
with asymptotic polynomial behavior,
that has already been proved super-renormalizable (only one-loop divergences survive), 
is actually completely regularized even at one-loop order using the Pauli-Villars regularization procedure introduced by 
Diaz-Troost-van Nieuwenhuizen-van Proeyen (DTPN)  
and applied to Einstein's gravity by Anselmi.
The regularization mechanism is diagrammatically shown in Fig.\ref{fig1}. 
Multi-loops amplitudes are finite because of the higher derivative scaling of the theory in the ultraviolet regime.
Since there are no divergent contributions, 
all the beta functions are zero.
 It turns out that 
 the theory is completely finite to any order in the loop expansion. 
 
\end{document}